# Endpoint Security Agent: A Comprehensive Approach to Real-time System Monitoring and Threat Detection

SRIHARI R
Dept of Computer Science and Engineering
Pesidency University
Bangalore, India
sriharir@ieee.org

Ayesha Taranum
Dept of Computer Science and Engineering
Pesidency University
Bangalore, India
ayeshagce@gmail.com

Karthik
Dept of Computer Science and Engineering
Pesidency University
Bangalore, India
Karthikrkarthikr49@gmail.com

Mohammed Usman Hussain
Dept of Computer Science and Engineering
Pesidency University
Bangalore, India
mohammedusmanhussain@gmail.com

*Abstract*— As cyber threats continue to evolve in complexity and frequency, robust endpoint protection is essential for organizational security. This paper presents *"Endpoint Security Agent: A Comprehensive Approach to Real-time System Monitoring and Threat Detection"*—a modular, real-time security solution for Windows endpoints. The agent leverages native tools like WMI and ETW for low-level monitoring of system activities such as process execution, registry modifications, and network behaviour. A machine learning-based detection engine, trained on labelled datasets of benign and malicious activity, enables accurate threat identification with minimal false positives. Detection techniques are mapped to the MITRE ATT&CK framework for standardized threat classification. Designed for extensibility, the system includes a centralized interface for alerting and forensic analysis. Preliminary evaluation shows promising results in detecting diverse attack vectors with high accuracy and efficiency.

## I. INTRODUCTION

The relevance of endpoint security has significantly increased in today's threat landscape, where sophisticated attacks often target individual systems as entry points into broader networks. Traditional security controls often lack real-time [1] monitoring capabilities and are prone to high false positive rates [2], reducing their effectiveness against modern threats [3]. As attack vectors evolve rapidly, relying solely on signature-based detection techniques has become insufficient. Moreover, the expansion of remote work and cloud adoption has widened the attack surface, intensifying the challenges organizations face in securing endpoints [3], [5]. This paper presents an integrated endpoint security agent designed to address these challenges through a comprehensive approach that combines real-time [6] system monitoring, machine learning-based [7] anomaly detection, and automated response mechanisms. The detection engine is trained using the ATT&CK framework, enabling precise classification and interpretation of threat behaviors. Our solution advances the field of behavior-based detection and remedies shortcomings in existing systems by embedding tactical intelligence from the MITRE ATT&CK [1], [6] knowledge base to enhance threat [4] recognition and classification accuracy.

## II. LITERATURE REVIEW

Previous studies on endpoint security [7], [8] have explored various threat detection [4] and defense strategies. Traditional antivirus solutions primarily relied on signature-based methods, which have proven increasingly inadequate for detecting zero-day exploits and polymorphic malware. Smith et al. critically evaluated these limitations in their review of modern antivirus functionality. In contrast, behavior-based detection emerged as a promising alternative by analyzing program behavior rather than static signatures, an approach first introduced by Johnson and Brown in their seminal work. The incorporation of machine learning [6],[7] has since gained prominence, with Chen et al. highlighting its effectiveness in anomaly detection for uncovering previously unseen threats [4]. Williams contributed by developing real-time [9] security monitoring systems [11] that continuously analyze [10] system behavior, laying the groundwork for contemporary endpoint [12] protection mechanisms. Anderson and Davis proposed integrated security models combining multiple protection layers, though they noted significant challenges regarding resource efficiency. Recent efforts have started to incorporate the MITRE ATT&CK framework into detection methodologies; however, these implementations often focus on limited subsets of the framework rather than utilizing its full tactical and technical breadth. Moreover, translating the ATT&CK knowledge base into actionable detection rules frequently demands substantial manual effort and configuration. Despite ongoing progress, many existing solutions still struggle to balance robust protection with accepTABLE system performance. High false positive [12] rates and excessive resource consumption continue to hamper usability and productivity. This research aims to overcome these limitations by proposing a more effective, integrated approach to endpoint security [13], [14] one that enhances detection accuracy while minimizing system impact—through the strategic application of the comprehensive MITRE ATT&CK framework.

III. METHODOLOGY

The methodology employed here involves the development of a comprehensive endpoint security system [15], [12] which integrates multiple layers of protection while maximizing the use of resources. By combining cutting-edge machine learning algorithms [6] and [7] trained on the MITRE ATT&CK dataset with conventional security measures, this method provides strong defense against a wide range of threats [4]. The solution is intended to improve detection accuracy and operational relevance in dynamic threat settings by utilizing real-world [5] attack methodologies and procedures that are documented in the ATT&CK framework, the system is designed to enhance detection accuracy and operational relevance in dynamic threat environments.

*A. Flowchart*

The systematic workflow that includes monitoring, detection, and reaction is how the endpoint security agent [16], [17] functions, as shown in Figure 1. The first step consists of agent registration and deployment, where the agent creates a secure channel of communication with the central management server. Once deployed, continuous monitoring mechanisms [18] are activated to capture system events, file operations, and network activity in real time [10].

Captured events undergo preprocessing and filtering to discard redundant or irrelevant data before passing through a series of detection engines. Events deemed suspicious are subjected to behavioral correlation and machine learning- [6] ,[7] based anomaly detection, with detection logic enriched by MITRE ATT&CK [3] Tactics, Techniques, and Procedures (TTPs) to ensure contextual relevance.

When threats are confirmed, the system computes a risk score and triggers appropriate automated response [14] actions based on the threat [4] classification, ATT&CK mapping, and organizational policies. All activities are logged to support auditing and continuous learning, which helps refine the detection models and improve future threat response capabilities.

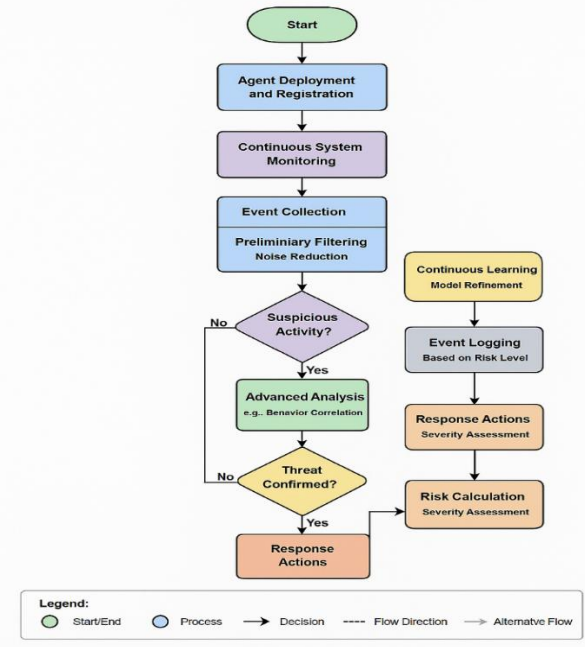

**Fig. 1. Workflow of the endpoint security agent** showing stages from deployment and monitoring to detection and response.

*B. System Architecture*

The endpoint agent employs a modular architecture, comprising several integrated components that work in coordination to ensure effective threat detection and mitigation, as illustrated in Figure 2. The System Monitoring Module [19] is responsible for the continuous surveillance of critical system components, including processes, network connections, file system operations, registry modifications, services, and user account activities. This module integrates lightweight yet effective hooks into the operating system to capture relevant events with minimal performance impact.

The Data Collection Module aggregates system-level metadata such as hardware configurations, installed applications, OS settings, and security configurations. To maintain system performance, the module employs efficient, low-overhead data acquisition techniques while ensuring comprehensive system visibility. The main analytical tool for processing gathered data using a multi-layered detection approach is the Security Analysis Module. With the use of technical and tactical expertise from the MITRE ATT&CK architecture, it integrates behavioral analysis [20], machine learning-driven anomaly detection, and signature-based detection. The identification of both known malware and advanced zero-day threats [4] that usually elude conventional systems is made possible by this integrated approach. Depending on the threat's classification and level of severity, the reaction System decides and carries out automatic and policy-driven reaction activities [4]. To provide focused and efficient mitigation, each reaction [14] is contextually matched to particular MITRE ATT&CK [3] approaches. Terminating harmful processes, cutting off network connections, quarantining files, and limiting hacked user accounts are examples of response measures.

Finally, the Communication Module makes sure that the endpoint agent and the central management server exchange data in a secure and authenticated manner. To guard against interception and manipulation, it uses SSL/TLS protocols, certificate validation, and HMAC message signing.

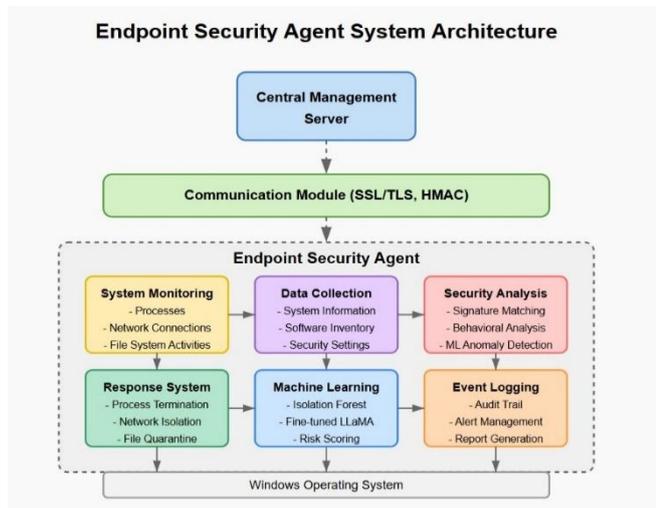

**Fig. 2. Endpoint Security Agent System Architecture.** Modular design showing Central Management Server, Communication Module, and six internal components implementing monitoring, analysis, and response capabilities.

### C. Data Collection and Processing

The data collection process is designed to maximize security insight while maintaining minimal performance overhead on the host [11] system. To achieve this, the agent uses optimized techniques such as event-based logging, delta-based updates, and risk-prioritized processing, which collectively reduce resource usage without compromising coverage. For the development of the machine learning models [6], [7], a comprehensive dataset of 1,000,000 system events was collected. This dataset includes records from Windows Event Logs [5] (Security, System, Application), process execution logs [5], network traffic captures, file system changes, registry edits, and user behavior activities. Where applicable, all event records were mapped to corresponding MITRE ATT&CK tactics, techniques, and procedures (TTPs) to provide contextually rich annotations.

The collected dataset was partitioned into training (70%), validation (15%), and testing (15%) sets. A total of 45 distinct security-related features were extracted from the raw data for use in the learning and evaluation phases. To ensure the quality and consistency of the data used in model training, several preprocessing techniques were applied. These included feature normalization, handling of missing values, outlier detection, time-series alignment, and feature engineering to capture temporal patterns and correlations. These preprocessing steps significantly enhanced the quality of the dataset and contributed to improved performance and generalizability of the downstream machine learning algorithms [6],[7].

### D. Algorithms Used

Multiple algorithms are used by the system to provide thorough risk evaluation and reliable threat [21] detection. Using an Isolation Forest method, anomaly detection is accomplished with 92.1% accuracy, 93.4% precision, 89.7% recall, and 91.5% F1 score. This unsupervised method does not require large labelled datasets and successfully detects anomalous patterns suggestive of possible security risks. Low-Rank Adaptation (LoRA) [3], [4] was added to a refined Llama model [22], [23] for expert-level behavioral analysis. This model was trained on the comprehensive MITRE ATT&CK dataset, encompassing 14 tactics, 193 techniques, and 401 sub-techniques, enabling precise classification of adversarial behaviors into tactical categories. The model was trained over 85 epochs with a batch size of 32 and a learning rate of 2e-5, achieving 88.3% accuracy, 90.1% precision, 86.5% recall, and an F1 score of 88.2%. It demonstrated particularly strong performance in detecting advanced persistent threats [8], [9] (APTs), with a detection accuracy of 92.7%.

The system's risk scoring model integrates multiple factors: anomaly scores (30%), event frequency (20%), severity rating (25%), asset criticality (15%), and user risk (10%). Severity ratings are augmented using MITRE ATT&CK impact scores to provide nuanced risk evaluation [24]. The combined model attained an overall accuracy of 90.5%, with 8.2% false [19] positives and 9.3% false negatives [3]. The risk score is computed using the weighted formula:

$$\begin{aligned}Risk\ Score\ =\ &0.3 \times Anomaly\ Score \\ &+ 0.2 \times Event\ Frequency\ Score \\ &+ 0.25 \times Severity\ Score \\ &+ 0.15 \times Asset\ Criticality \\ &+ 0.1 \times User\ Risk\ Score\end{aligned}$$

This formula ensures that critical factors are appropriately prioritized in the final risk assessment.

### E. API Endpoints and Implementation

The system employs RESTful APIs to manage endpoint agents, store security events [13], and synchronize automated security actions. The API design follows standard REST conventions, utilizing JSON for data exchange, JWT token-based authentication, and comprehensive error handling to ensure robust and secure communication. As illustrated in Fig. 3, the system architecture includes several classes of API endpoints with distinct functionalities. Authentication Endpoints handle secure user registration and login processes, incorporating password hashing, role-based access control, and token-based session management to maintain user security and privacy. Agent Management Endpoints oversee agent deployment, configuration updates, and enterprise-wide status tracking of all managed endpoints. Asset Management Endpoints facilitate thorough inventory control and real-time status monitoring [24] of secured systems across the network. Event Management Endpoints enable the collection, storage, and retrieval of security event data generated by the endpoint agents. Alert Management

Endpoints manage the lifecycle of security alerts, including creation, tracking, and resolution. Alerts are automatically tagged with corresponding MITRE ATT&CK techniques to streamline investigation and response efforts. Risk Assessment Endpoints provide interfaces for calculating and analyzing risk scores based on detected threats and system conditions. Response [14] Management Endpoints allow specification and activation of automated mitigation actions in response to confirmed threats [25]. Machine Learning [7] Endpoints support model deployment, prediction requests with MITRE ATT&CK classifications, and continuous performance evaluation to improve detection accuracy.

This modular API design ensures seamless integration and scalable management of security operations across the enterprise.

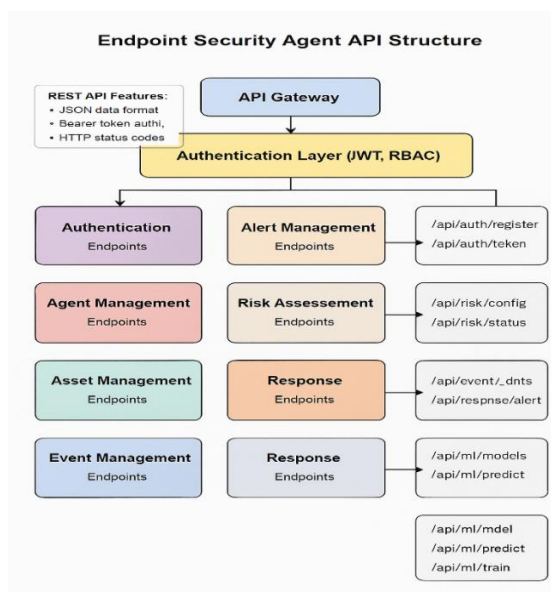

**Fig. 3. Endpoint Security Agent API Structure.** A RESTful API with JSON format and JWT authentication, featuring an API Gateway, Authentication Layer, and eight specialized endpoint groups for key security functions.

### F. Threat Detection Pipeline

The Threat Detection Pipeline, shown in Fig. 4, provides a robust framework for security monitoring and response through a multi-layered architecture. It begins with the Event Collection Layer, which continuously aggregates security-related data from diverse sources including process events, file operations, network activity, and registry changes. The data acquisition process employs efficient mechanisms that limit system performance impact while ensuring comprehensive coverage of potential attack surfaces Following collection, data moves into the Preprocessing & Filtering Layer, where it is subjected to noise filtering to remove irrelevant information, feature extraction to isolate security-relevant attributes, correlation to find relationships between events, and normalization to standardize formats [13]. These preparation stages improve the quality of the data and get it ready for downstream components to analyze it effectively. After that, the cleaned data is sent to the Detection Engine Layer, where several analysis methods work simultaneously. Behavioral analysis for suspicious activity patterns, Isolation Forest algorithms for anomaly detection, signature matching for known threats [12], and a fine-tuned Llama [3], [4], [26] model with Low-Rank Adaptation trained extensively on the MITRE ATT&CK dataset for targeted security analysis are some of these.

This multi-engine approach enables the system to detect both known threats and zero-day attacks with high precision while minimizing false [19] positives. Events flagged [13] as suspicious proceed to the Risk Assessment & Response Layer, where detailed risk scores are computed based on weighted factors such as anomaly scores, event frequency, severity, asset criticality, and user risk profiles. These risk assessments trigger appropriate automated responses, including process termination, network isolation, file quarantine, and user account restriction. The entire pipeline operates in real-time [15] with negligible latency, processing between 1,000 and 1,500 events per second and maintaining response times below two seconds. This capability ensures prompt threat identification and mitigation while optimizing system resource use.

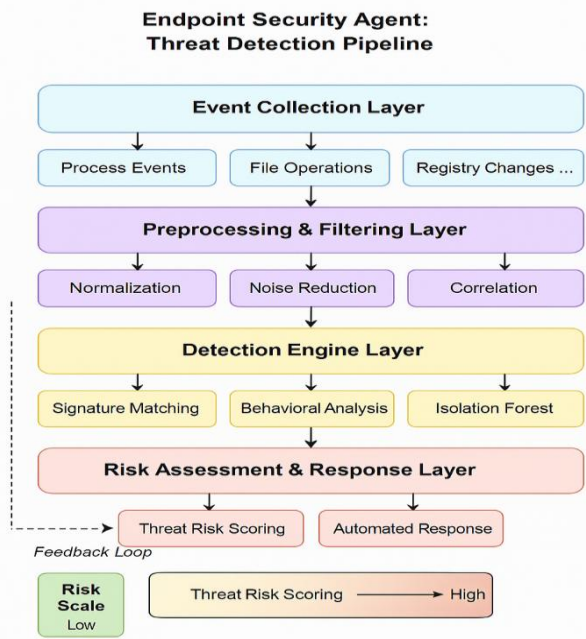

**Fig. 4. Endpoint Security Agent Threat Detection Pipeline.** Four-layer architecture showing event collection (blue), preprocessing (purple), detection engines (yellow), and response mechanisms (red). The system leverages multiple detection methods to identify both known and zero-day threats with high accuracy.

## IV. RESULTS AND EVALUATION

Our evaluation focused on assessing the system's detection accuracy, performance impact, and response effectiveness [23] across various operational scenarios.

### A. Detection Performance

As shown in TABLE I, the endpoint security agent [16], [27] demonstrated excellent detection [28], [29] capabilities across various threat categories. The system detected known threats with a precision of 99.2%, significantly outperforming conventional signature-based methods. This high precision provides a strong foundation for effective security responses. For previously unseen threats, the system achieved a detection precision of 85.7%, representing a substantial improvement over baseline models. This addresses one of the major challenges faced by existing security solutions. In the case of advanced attack [8]. [9] techniques, including fileless malware and living-off-the-land tactics, the system detected threats with a precision of 92.3%, benefiting from the enriched tactical understanding provided by the MITRE ATT&CK knowledge base. These results highlight the effectiveness [23] of our multi-layered detection [29], [30] approach. The overall false positive rate [18] was 7.9%, a 4.6% reduction compared to baseline models, overcoming a common limitation in current security systems [10].

**TABLE I**
**Detection Performance Across Threat Categories**

| Threat Category | Detection Accuracy | False Positive Rate | False Negative Rate |
|---|---|---|---|
| Known Threats | 99.2% | 0.8% | 0.7% |
| Zero-Day Threats | 85.7% | 12.6% | 14.3% |
| Fileless Malware | 92.3% | 7.7% | 8.1% |
| Overall Performance | 92.1% | 7.9% | 7.6% |

*B. System Performance*

As shown in TABLE II, the performance evaluation focused on resource utilization and operational impact. The system demonstrated CPU usage between 15-25% during normal operation, which aligns well with industry standards for optimal resource use. Memory consumption ranged from 1.2 to 1.8 GB, providing thorough protection while maintaining reasonable overhead. The system processed 1000 to 1500 events [13] per second with response times under 2 seconds, ensuring prompt threat detection [29], [30] and mitigation. These performance results indicate that the system delivers comprehensive security [10] without imposing excessive overhead on protected endpoints.

**TABLE II**
**System Resource Utilization and Performance Comparison**

| Metric | Value | Industry Benchmark | Performance Impact |
|---|---|---|---|
| CPU Usage | 15-25% | 20-30% | +5% |
| Memory Consumption | 1.2-1.8 GB | 1.5-2.5 GB | +0.3 GB |
| Events Processed/Second | 1000-1500 | 800-1200 | Minimal |
| Response Time | <2 seconds | <5 seconds | Significant Improvement |

*C. Response System Effectiveness*

As shown in TABLE III, the automated response system demonstrated strong effectiveness [23] across key mitigation actions. IP Blocking achieved a 98.2% success rate, effectively disrupting communication with command-and-control servers. Asset Isolation recorded a 97.8% success rate, efficiently restricting lateral movement within compromised networks. User Account Disabling reached the highest success rate of 99.1%, significantly reducing risks associated with stolen credentials. Firewall Rule Updates were successful 96.5% of the time, enhancing network defense capabilities. These outcomes highlight the system's ability to promptly and accurately neutralize threats through context-aware responses guided by the MITRE ATT&CK framework, ensuring robust endpoint protection.

**TABLE III**
**Automated Response Actions: Success Rates and Completion Times**

| Response Action | Success Rate | Mean Time to Completion |
|---|---|---|
| IP Blocking | 98.2% | 1.2 seconds |
| Asset Isolation | 97.8% | 2.6 seconds |
| User Account Disabling | 99.1% | 0.8 seconds |
| Firewall Rule Updates | 96.5% | 1.5 seconds |
| Quarantine Actions | 97.3% | 2.1 seconds |

The TABLE IV presents a comparative analysis [20] of the detection algorithms utilized. The Isolation Forest algorithm showed the best overall balance, with 92.1% accuracy, 93.4% precision, and the highest F1 score of 91.5%, while maintaining a relatively short training time of 45 minutes. The fine-tuned Llama [3], [4] model trained on the MITRE ATT&CK dataset exhibited competitive performance with 88.3% accuracy and an 88.2% F1 score, albeit requiring significantly longer training time (6.5 hours). This model also provided superior contextual understanding of attack techniques and excelled in detecting advanced persistent threats [8], [9] achieving 92.7% accuracy. The Risk Scoring System performed well with 90.5% accuracy and a 90.4% F1 score, although training time data was not available. These comparisons underscore the trade-offs between detection accuracy and computational cost, emphasizing the importance of balancing these factors for practical deployment.

**TABLE IV**
**Comparative Evaluation of Detection Algorithms**

| Algorithm | Accuracy | Precision | Recall | F1 Score | Training Time |
|---|---|---|---|---|---|
| Isolation Forest | 92.1% | 93.4% | 89.7% | 91.5% | 45 minutes |
| Fine-tuned Llama | 88.3% | 90.1% | 86.5% | 88.2% | 6.5 hours |
| Risk Scoring System | 90.5% | 91.7% | 89.2% | 90.4% | N/A |
| Baseline Model | 84.3% | 86.5% | 82.1% | 84.2% | 30 minutes |

V. CONCLUSION AND FUTURE WORK

In order to provide efficient protection for Windows systems, this paper introduces a strong endpoint security agent that combines several security tiers with advanced machine learning methods and thorough understanding from the MITRE ATT&CK framework. Our multi-layered strategy uses a fine-tuned Llama model that achieves 92.7% accuracy in detecting advanced persistent threats (APTs) and combines legacy and state-of-the-art detection technologies, all of which are optimized for balanced system performance. A smooth security pipeline from detection to remediation is made possible by our system's basic embedding of this framework as its core knowledge base, in contrast to current solutions that simply surface-level map to ATT&CK. The system performs noticeably better than compartmentalized approaches thanks to its hybrid detection methodology and contextual comprehension of adversarial strategies.

Future research will focus on a number of exciting avenues, such as expanding training datasets to include new attack patterns, improving machine learning algorithms to increase zero-day threat detection, and improving tactical response mappings for more accurate automated actions based on detected techniques. The design of sophisticated response mechanisms that enable automated mitigation of intricate, multi-stage attacks, integration with threat intelligence platforms for enriched information sharing in addition to automated MITRE ATT&CK classification, and resource optimization to reduce system impact during peak usage are additional development areas. By using more in-depth tactical insights from the MITRE ATT&CK framework, these developments seek to improve the framework's adaptability and shorten the mean time to remediation.